# Noncontact electrical metrology of Cu/low-*k* interconnect for semiconductor production wafers

Vladimir V. Talanov, André Scherz, and Andrew R. Schwartz

*Neocera, Inc., 10000 Virginia Manor Road, Beltsville, MD 20705, USA*

We have demonstrated a technique capable of *in-line* measurement of dielectric constant of low-*k* interconnect films on patterned wafers utilizing a test key of ~50×50 μm$^2$ in size. The test key consists of a low-*k* film backed by a Cu grid with >50% metal pattern density and <0.25 μm pitch, which is fully compatible with the existing dual-damascene interconnect manufacturing processes. The technique is based on a near-field scanned microwave probe and is noncontact, noninvasive, and requires no electrical contact to or grounding of the wafer under test. It yields <0.3% precision and ±2% accuracy for the film dielectric constant.

As the shortening integrated circuit (IC) development circles are pushing the semiconductor industry to employ manufacturing processes that are not as mature as once possible [1], the International Technology Roadmap for Semiconductors (ITRS) is calling for use of the *in-line* and/or *in-situ* metrology techniques during all phases of semiconductor manufacturing – process research and development, yield ramp, and volume production [2]. However, there is a distinct lack of such metrologies capable of electrical evaluation of modern Cu/low-*k* interconnect – the structure formed by up to a dozen levels of Cu wiring separated by interlevel dielectric (ILD), which constitutes more than half of the IC fabrication cost. Porous low-*k* dielectrics aimed to even further reduce the delay times, crosstalk and power consumption of interconnect pose significant integration and fabrication challenges, such as variation across the wafer of the dielectric constant, *k*, which is sensitive to physical and chemical damage caused by various damascene processing steps (i.e. the process in which Cu is inlaid in the trenches and vias pre-etched in ILD) [3, 4]. The goal of this paper is to demonstrate a metrology technique addressing these issues on a semiconductor *production* wafer.

Neither of the three methods commonly in use today for electrical evaluation of *blanket* interconnect stacks, namely mercury probe [3], metal-insulator-semiconductor capacitor (MIS-cap) [3], and Corona discharge/Kelvin probe [5], can be used on patterned wafers. While the first two are employed *off-line* only, the third can be applied *in-line* to blanket *monitor* wafers. The interdigital or comb capacitor [3] is routinely implemented on *patterned* wafers and is more representative of real interconnect properties. However, it requires significant space, is time consuming (patterning, extensive cross-sectioning and numerical modeling are involved) and fairly inaccurate due to fringing effects and statistical pattern variation [3, 6]. Recently, NIST researchers developed a microstrip transmission line technique to show that up to at least 40 GHz low-*k* materials exhibit low dielectric losses and little dispersion [7]. While this is the only technique capable of measuring the *complex* permittivity, it too is destructive and time/labor intensive (patterning, cross-sectioning and modeling are required). Dielectric constant measurements of blanket low-*k* films with an evanescent microwave probe have been reported [8], but the use of a 100-micron tip in physical contact with the sample precludes this technique from being applied on production wafers. Thus, there is an unmet need for a metrology capable of *in-line* monitoring of the ILD dielectric constant within the 80-µm-wide scribe line or active die region *on production* wafers. The primary requirement for such a technique is that the probe sampling spot size must be <50 μm. The metrology should be non-destructive, non-contaminating, and provide real time data collection and analysis on a test key that is compatible with the existing damascene process. We have developed such a method based on our near-field scanned microwave probe (NSMP) featuring noncontact operation and a sampling spot size of about 10 μm [9].

In the near-field approach the spatial resolution is governed by the probe tip geometry rather than the wavelength of the radiation utilized. Therefore, even at microwave frequencies with wavelengths on the order of centimeters, sub-micron spatial resolution is possible [10]. Our apparatus and the method for quantitative measurements on *blanket* low-k wafers have been described elsewhere [9, 11]. The near-field probe is a half-lambda parallel strip transmission line resonator (PSR) microfabricated from a quartz bar tapered down to a few micron tip size and sandwiched between two aluminum strips (Fig.1). The PSR is mounted inside a metallic sheath with the tip protruding out via an opening in the sheath wall and operates in a 4 GHz balanced odd mode. The near-zone field is mostly confined in between the Al strips. The tip sampling *E*-field (similar to the fringe field of a parallel plate capacitor) forms a well-confined "cloud" with a characteristic dimension on the order of the tip size $D$~5–10 μm. When a dielectric sample is brought in close proximity to the tip, the reactive energy stored in this field is reduced, and consequently the probe resonant frequency $F$ decreases.

Since the tip is much smaller than the radiation wavelength, a lumped element network can be used to describe the tip-sample interaction. For a low-loss dielectric film the tip impedance $Z_t$ can be represented as a network of the air-gap capacitance $C_g$, the film capacitance $C_f$, and the backing impedance $Z_b$ (Fig.1). If $|Z_b|<<1/\omega C_f$, $1/\omega C_g$ then $Z_t = 1/i\omega C_t = 2(C_f^{-1} + C_g^{-1})/i\omega$ and the relative shift of the probe resonant frequency, $F=\omega/2\pi$, versus change in the tip capacitance $C_t$ is [11]:

$$\Delta F/F_0 = -2F_0 Z_0 \Delta C_t \tag{1}$$

where $Z_0 \approx 100\Omega$ is the characteristic impedance of the transmission line, $F_0$ is the probe frequency without sample present, and $\omega Z_0 C_t <<1$. The tip-sample distance is controlled by a shear-force feedback method [9] providing $C_g \geq C_f$, which ensures the sensitivity to film permittivity. For our frequency measurement precision ~0.2 ppm Eq. (1) predicts a capacitance sensitivity of ~0.3 aF, which is about six orders of magnitude better than that for Hg-probe or

MIS-cap (e.g. >0.1 pF). Therefore, if $C_f$ is a parallel plate capacitor formed by a typical 0.4-μm-thick low-$k$ film with $k$~2 its area needs to be only ~10 μm$^2$ to measure $k$ with 0.1% precision. By comparison, the area required for similar measurement by Hg-probe or MIS-cap is ~1 mm$^2$. The measured $\Delta F$ is converted into film $k$-value (with precision 1σ<0.3% and accuracy better than ±2%) using the approach developed in Ref. [11], which employs for calibration a set of variable thickness SiO$_2$ films on low resistivity Si substrates with $|Z_b|<<1/\omega C_f$. Wafers up to 300 mm in diameter can be scanned. The apparatus is equipped with optical navigation and sits on a vibration-isolated platform inside an environmental chamber at ambient conditions.

The proposed test key for the measurement is a ~50×50 μm$^2$ open area formed by a low-$k$ film (with or without hard-mask and/or etch-stop layers) backed by either continuous or patterned metal (e.g. Cu) as will be discussed below. The backing should provide $|Z_b|<<1/\omega C_f$ to conform to the calibration method, while also shielding the underlying interconnect levels to avoid their contribution into the response. The test key is compatible with damascene processing, can be repeated throughout all interconnect levels, and fits easily into an 80-μm-wide scribe line between die.

To confirm that the measurement is unaffected by the finite size of the test key, we measured the probe frequency $F$ vs. lateral position across the edge of a 60×60 μm$^2$ Cu patch buried under a 414-nm-thick low-$k$ film (the film thickness was measured by ellipsometry at the same site). The structure was located on a patterned 300 mm wafer and surrounded by other test structures (Fig. 2, site A). Figure 3 shows the extracted $k$-value as a function of position $x$ normalized by the tip size $D$~10 μm, where the Al strips forming the probe were oriented parallel to the patch edge (the scan with the Al strips perpendicular to the edge looks the same). One can see that outside of a transition range ~1.5×$D$ wide the "apparent" $k$-value (i.e. as seen by the probe) exhibits no change (within the measurement precision). Over the patch we obtained $k$=3.21±0.01, which is in agreement with the pristine value of 3.20. Note, if the measurement was affected by the finite patch size it would yield a *lower* than expected dielectric constant. Outside the patch the "apparent" $k$ exhibits a non-physical value ≤1 because in the analysis it was assumed that the film is 414-nm-thick, which is correct only above the patch. Assuming the worst possible case of the entire edge transition occurring *above* the patch, the maximal size of the test key is 2×(1.5×D)+D =4×D≈40 µm in order to accommodate the two transitions plus a single sampling spot size ~D.

In practice, the continuous Cu backing is not compatible with damascene processing because chemical mechanical planarization (CMP) technology used in fabrication of Cu interconnect is sensitive to pattern density (e.g., Cu lines wider than a few micron are not acceptable). Therefore, we investigated if a Cu grid could be used instead. Figure 4 shows the *k*-values measured on a set of 60×60 µm$^2$ test keys consisting of the same low-*k* film now backed by parallel Cu lines of variable pitch and pattern density (see Fig.2, sites B). These are actually optical scatterometry test keys (OCD), each with a fixed pitch and density. The Al strips forming the probe were oriented perpendicular to the lines to minimize $Z_b$. One can see in Fig. 4 that for pattern density of 50% and nominal pitch less than ~0.25 µm the data analysis based on a uniform backing with $|Z_b|<<1/\omega C_f$ yields the "apparent" *k*=3.14 independent of pitch and in good agreement with that observed on the continuous Cu patch. As the pitch increases and/or Cu density decreases, the "apparent" dielectric constant decreases because of reduction in the film geometrical capacitance due to change of the backing geometry. Thus, the grid backing with small enough pitch and high enough metal density yields the same "apparent" *k*-value as the continuous one, while making the result insensitive to inevitable statistical variations in the grid pattern. Note that for Cu density greater than 50% a larger pitch is probably acceptable, but such test keys were not available on the wafer under study. According to the lumped element scheme in Fig. 1, if the film thickness increases and/or *k*-value decreases the acceptable pitch/Cu density are expected to increase/decrease, and vice versa. Finally, it was observed that the patterned backing actually provides better site-to-site repeatability than the continuous one, likely due to improved Cu morphology.

To conclude, we have demonstrated that the dielectric constant of low-*k* ILDs can be quantitatively measured *in-line* on production wafers, even when the film is backed by patterned metal lines. This metrology has the following advantages: a) it is a noncontact, noninvasive, and noncontaminating (the probe is fabricated entirely from quartz and aluminum); b) the test key is fully compatible with the existing interconnect technology (dual-damascene and CMP); c) the test key is already in use by other metrologies such as ellipsometry and OCD/scatterometry [4], allowing for film thickness, *k*-value, and critical dimension (CD) measurements at the same site on the wafer; and d) no electrical contact to or grounding of the wafer is required since both probing electrodes are located above and capacitively coupled to it allowing for the test key to be placed at any level in the interconnect structure. The technique can be used for mapping of the *k*-value variation across the wafer after deposition, controlling the porogen removal, and characterization of the dielectric damage in low-*k* patterns.

This work was partially supported by NIST-ATP Award No. 70NANB2H3005. We are thankful to J. Matthews and Prof. T. Venky Venkatesan for stimulating discussions, and J. S. Tsai for providing the wafers used in this study.

## Figure captions

### Figure 1.

Apparatus schematic showing the probe and electronics. Inset: the lumped element scheme for the tip capacitance in the case of a low-$k$ film with arbitrary backing.

### Figure 2.

Optical photograph of the region on a 300 mm wafer with OCD test keys. Site A is a 60×60 μm$^2$ open area with a 414-nm-thick low-$k$ film backed by a continuous Cu patch. Sites B are 60×60 μm$^2$ open areas with the low-$k$ film backed by patterned Cu lines with variable pitch and density.

### Figure 3.

"Apparent" film dielectric constant vs. normalized position $x/D$ across the edge of a 60×60 μm$^2$ Cu patch located at $x>0$ and buried under a low-$k$ film with pristine $k=3.20$ (site A, Fig. 2); solid line is a guide to the eye. Tip size $D\sim10$ μm.

### Figure 4.

"Apparent" film dielectric constant vs. pitch of the backing formed by Cu lines embedded into low-$k$ material for four Cu pattern densities (sites B, Fig. 2); solid lines are guides to the eye.

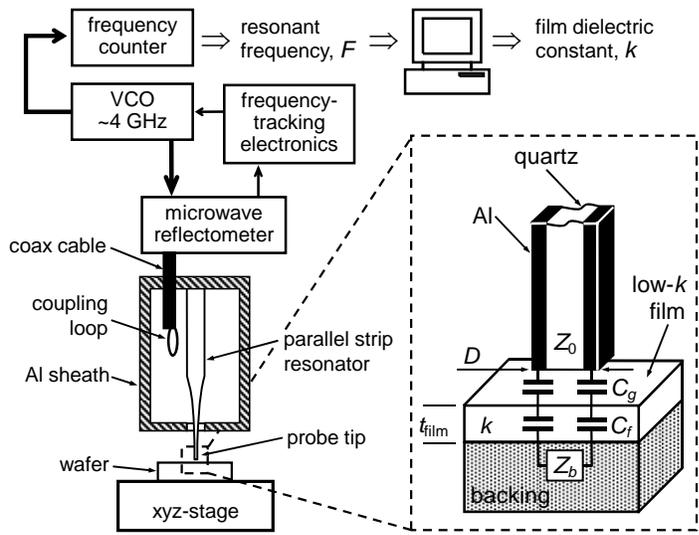

**Figure 1.**

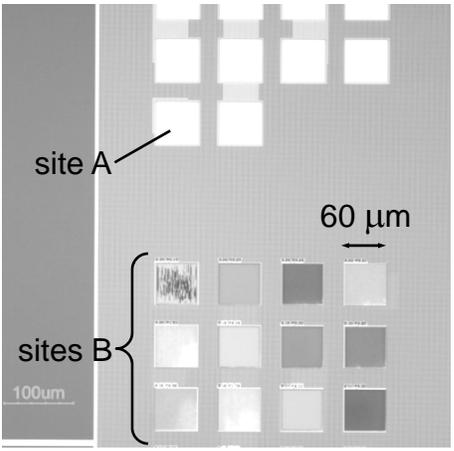

**Figure 2.**

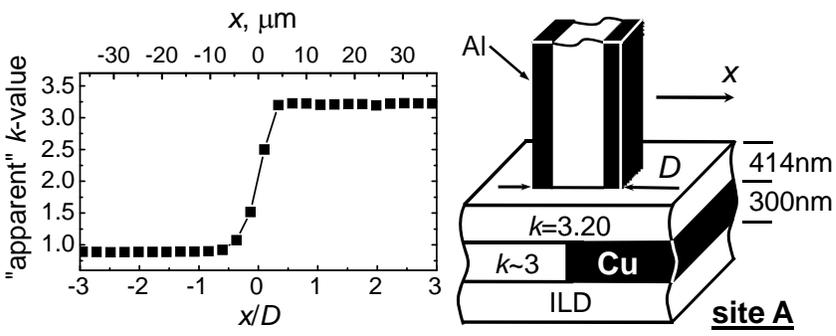

**Figure 3.**

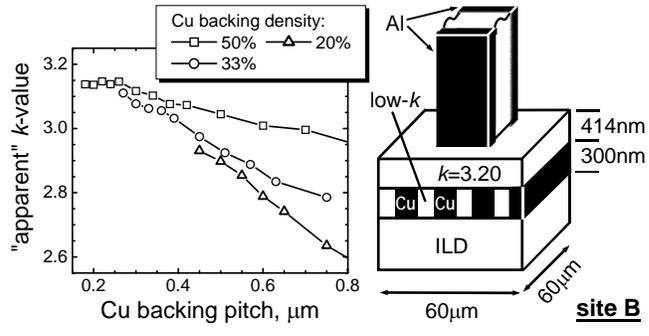

**Figure 4.**